\documentclass{iaus}
\usepackage{graphicx}
\def\rd{$r_d$}
\newcommand{\NII}{[{\sc N$\,$ii}]}
\newcommand{\OIII}{[{\sc O$\,$iii}]}

\def\kms{$\mbox{km s}^{-1}$}

\def\farcs{\hbox{$.\!\!^{\prime\prime}$}}
\def\arcsec{\hbox{$^{\prime\prime}$}}
\def\sun{\hbox{$\odot$}}

\title[A method to resolve the nuclear activity in galaxies] 
{A method to resolve the nuclear activity in galaxies, as applied to NGC~1358}

\author[Lindblad \& Fathi]   
{Per-Olof Lindblad$^1$ \& Kambiz Fathi$^1$}

\affiliation{$^1$Stockholm Observatory, Department of Astronomy, Stockholm University, Sweden\\email: {\tt kambiz@astro.su.se}}

\pubyear{2011}
\volume{277}  
\pagerange{\ }
\jname{Tracing the Ancestry of Galaxies on the Land of our Ancestors}
\editors{Editors: C. Carignan, F. Combes, K. C. Freeman}
\begin{document}

\maketitle

\begin{abstract}
Nuclear regions of galaxies generally host a mixture of
components with different exitation, composition, and kinematics.
Derivation of emission line ratios and kinematics could then be
misleading, if due correction is not made for the limited spatial
and spectral resolutions of the observations.
The aim of this paper is to demonstrate, with application to a
long slit spectrum of the Seyfert~2 galaxy NGC~1358, how line
intensities and velocities, together with modelling and knowledge
of the point spread function, may be used to resolve the differing
structures.
In the situation outlined, 
the observed kinematics differs for different spectral lines.
From the observed intensity and velocity distributions of a
number of spectral lines and with some reasonable assumptions
to diminish the number of free parameters, the true line ratios
and velocity structures may be deduced.
A preliminary solution for the nuclear structure of NGC~1358
is obtained, involving a nuclear point source and an emerging
outflow of high excitation with a post shock cloud, as well as a nuclear emission
line disk rotating in the potential of a stellar bulge and 
expressing a radial exitation gradient.
The method results in a likely scenario for the nuclear structure
of NGC~1358. For definitive results an
extrapolation of the method to two dimensions combined with the 
use of integral field spectroscopy will generally be necessary.

\end{abstract}

\firstsection 
\section{Overview}
The nuclear and circumnuclear activity of galaxies generally involves the
interplay between a number of different components and phenomena, e.g. a
central active source surrounded by an absorbing torus, a rotating central
bulge, outflowing jets, flowing streams due to the action of a bar, or even
merging. To separate these different components and derive their respective
line ratios and kinematic behaviour is generally difficult due to the limited
spatial and spectroscopic resolution available. On the other hand, such a
separation is crucial to the analysis of the structures and physical processes
involved in the nuclear region and their roles in galaxy evolution.
Then, as to be demonstrated here, such a separation could benefit by considering the
differences of the velocities observed for different spectral lines and be eased
by models of the activity, smoothed with the point spread function (PSF) and fitted
to the observations. Evidently, to yield the best results this should require an
integral field spectrum obtained at the best available spatial and spectral resolution.

NGC~1358 is a barred Sa galaxy hosting an active
galactic nucleus at a heliocentric velocity of 
$\approx$4100~\kms, included in the 
sample of Ulvestad \& Wilson (1989) as a Seyfert~2. 
It attracted our attention because
of its remarkable circumnuclear kinematics 
(Dumas et al. 2007, Lindblad et al. 2010).

\section{Modelling}
An explanation for the differences in velocity behaviour between the various emission lines would be that the whole central region consists of different gaseous components with different velocity behaviour, different excitation, and different dust absorption, and thus different sets of line ratios. Smoothed by the PSF, different distortions of the velocity field for the different lines will result. Assume that the region considered covers $N$ pixels and contains $K$ separate structural components, each with its own line ratios and set of velocities. If $J_i^{(\lambda)}$ and $V_i^{(\lambda)}$ are the observed line intensities and velocities at pixel number $i$ of a line with wavelength $\lambda$, we have

\[J_i^{(\lambda)} \, = \, \sum_j \sum_k \, I_{kj}^{(\lambda)} \, p(i-j) \;\mbox{\ \ and \ \ } \;
V_i^{(\lambda)} \, = \, \frac{1}{J_i^{(\lambda)}}\displaystyle{\sum_j \sum_k \, v_{kj} \, I_{kj}^{(\lambda)}  \, p(i-j)},
\]
where $I_{kj}^{(\lambda)}$ is the intrinsic intensity, in the line $\lambda$, of the component $k=(1, \dots, K)$ at the pixel $j$, and $v_{kj}$ the corresponding velocity, which is supposed independent of $\lambda$, and $p(x)$ the PSF. If $L$ is the number of lines measured, we have in the general case $KN(L+1)$ unknown and $2NL$ equations. In the simpliest case, where we only consider one component $(K=1)$, we have enough equations to solve for the unknown intensities and velocities for each $\lambda$, which can be made by CLEAN-like algorithms (H\"ogbom 1974). Further details and application to the spectra from NGC~1358 are given in Lindblad et al. (2010). The model has the following features:

\begin{itemize}
\item[a)] A spherical stellar bulge with central position coinciding with that of the continuum maximum, being the sum of two exponential spheres with mass distributions computed using the relation (2--170) of Binney \& Tremaine (1987). The spheres have scale lengths 0\farcs7 and 4\farcs0 respectively, and a mass ratio of 4:1. We have, somewhat ad hoc, assumed the intensities in the emission line disk to be the sum of two exponential disks with the same scale lengths as the bulge (here called Disk~1 and Disk~2), thus representing the inner and outer part of the total emission line disk.
\item[b)] A central emission line point source coinciding with the centra of the bulge and the emission line disk and at rest with respect to that centre. The redshift of this centre is a free parameter.
\item[c)] A gas flow, here called the "jet", seen from the nucleus out to 5 pixels (1\farcs3) from the nucleus with constant line of sight velocity $- 214$ \kms\ with respect to the point source, which was the velocity of the "fainter" component relative to the "stronger" as seen in high spatial resolution spectra from the HST of the very nucleus.
\item[d)] High positive velocities are necessary to reproduce the sharp rise of the rotation curves at the end of the jet around +2\arcsec\ from the centre. We call this apparently post shock feature the "Cloud".
\end{itemize}
The resulting intensities and velocities are convolved with the observed PSF.The free parameters are then varied until satisfactory fits to the observed velocities and intensities have been obtained (see Lindblad et al. 2010). 

\begin{figure}
\begin{center}
 \includegraphics[width=.85\textwidth]{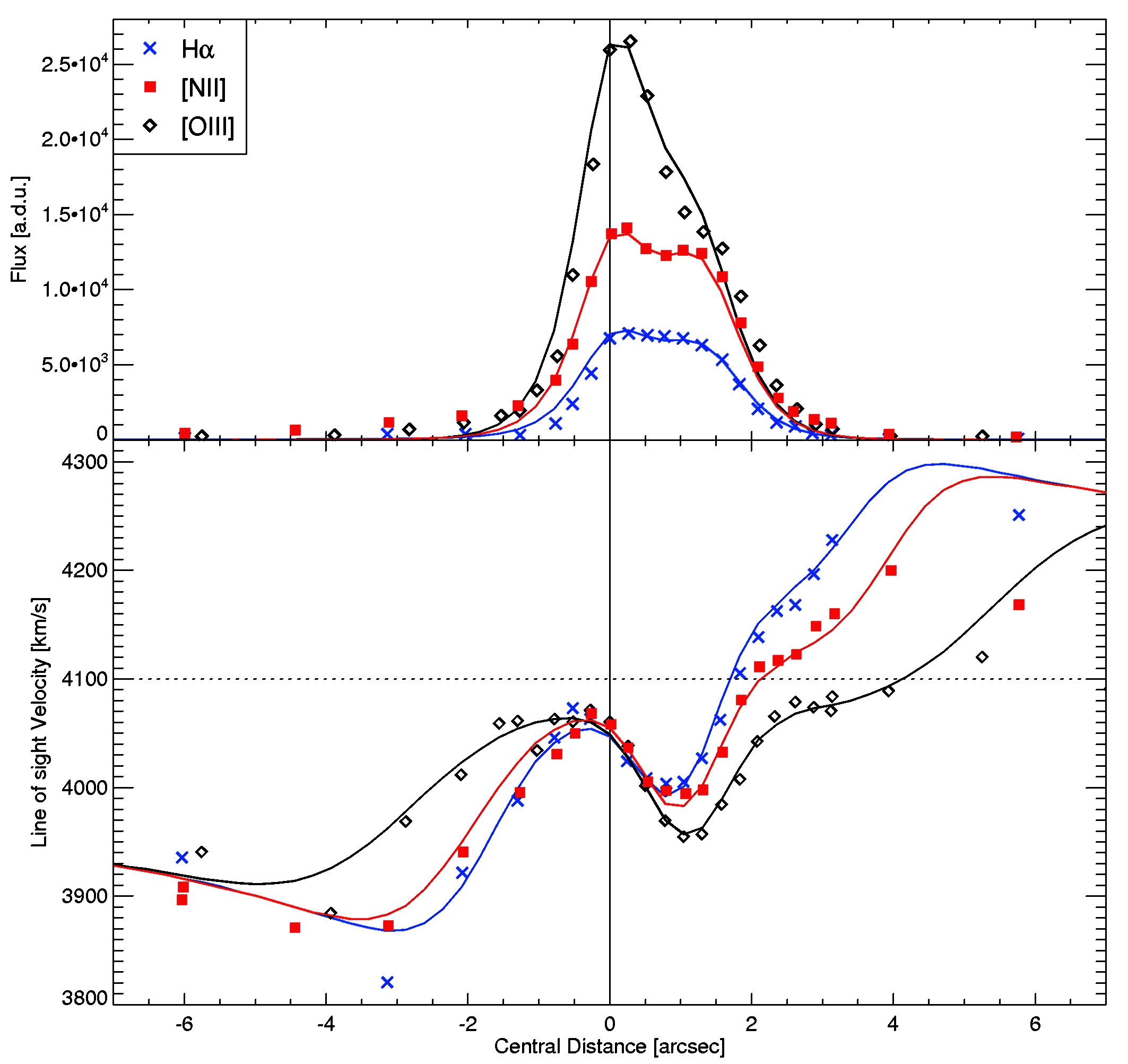}
  \caption{Best model fit to the emission line intensities and line of sight velocities. The dotted line indicates the derived redshift of the nuclear source. East is to the right.} 
\label{fig:model}
\end{center}
\end{figure}

\section{Conclusion}
Our model (Fig.~\ref{fig:model}) resolves the nuclear structure of NGC~1358 into (1) a central unresolved emission line source (the Nucleus), (2) a 1\farcs3 long jet emerging from the nucleus with a line of sight velocity of $-214$ \kms, (3) a spherical nuclear stellar bulge containing a rotating emission line disk inclined to the stellar kinematic symmetry plane of the bulge, (4) an emission line region (the Cloud) outside the jet with positive velocity as the emission line disk. There is no room for and no need for a counter-jet in our model. Then the true nuclear redshift must be higher than the observed one. This best fit gives an observed redshift of the central source of 4100 \kms, or 4089 \kms\ with heliocentric velocity correction. The total $M \times \sin^{2}(i)$ of the bulge is determined essentially by the velocities of H$\alpha$ and \NII\ at the distance $-6$\arcsec from the centre. The best fit gives $M \times \sin^{2}(i) = 14 \times 10^{9}~\rm M_{\sun}$. Figure~\ref{fig:model} shows that the velocities and intensities have been reproduced with fair accuracy within this nuclear region. The H$\beta$ intensities are too weak and uncertain to give a unique solution in the present case.

\begin{figure}
\begin{center}
 \includegraphics[width=.49\textwidth]{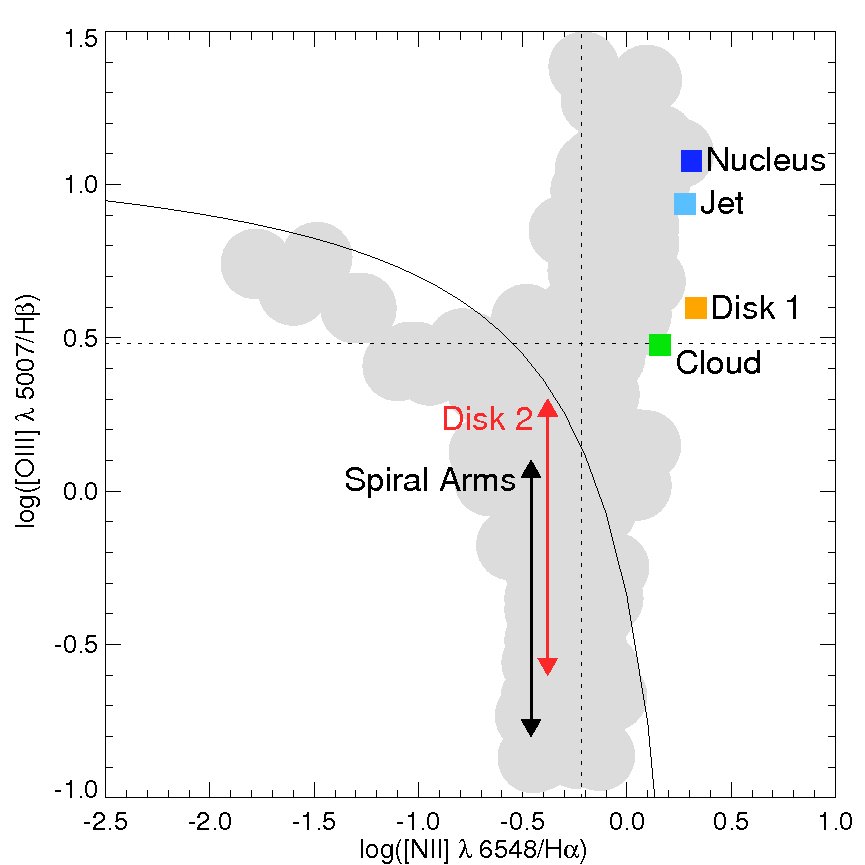}
  \caption{Line ratio diagnostics derived from our ESO spectra. The solid curve shows the Starburst-AGN separation lines of Kewley et al. (2001), with the shaded region indicating their sample galaxies.} 
\label{fig:lineratios}
\end{center}
\end{figure}

With the emission line fluxes for these structures we are able to set up the Starburst-AGN diagnostic diagrams introduced by Baldwin, Phillips \& Terlevich (1981). Figure~\ref{fig:lineratios} shows the diagnostic diagram \OIII/H$\beta$ versus \NII/H$\alpha$ where the \OIII/H$\beta$ ratio is derived from \OIII/H$\alpha$ assuming large optical depth, referred to as Case B described in chapter 4.2 of Osterbrock (1989). Accordingly, for an electron temperature of 10000 K, the H$\alpha$/H$\beta$ is 2.87, which we note is not much different from the case when the gas is assumed to be optically thin. As evident, the nuclear source and the jet fall in the region of highly excited AGNs, while the inner disk and the cloud fall close to the region of liners. The ratios \NII/H$\alpha$ places the outer part of the nuclear disk and the spiral arms in the HII region domain.

The contribution from the underlying H$\alpha$ and H$\beta$ absorption lines only change the derived emission line ratios marginally. However, the effect of reddening has not been taken into account in the emission ratios presented in Fig.~\ref{fig:lineratios}. As we have used the unreddened ratio H$\alpha$/H$\beta$, our derived \OIII/H$\beta$ ratios presented in the diagram are lower limits in the presence of reddening. This is an effect that favours our conclusions about the nature of the components (further details in Lindblad et al. 2010).

We argue that, with this analysis, we have made a step towards a resolution of the nuclear region of NGC~1358 into a number of different components with different velocity behaviour and excitation, where a summary of the results are seen in Fig.~\ref{fig:lineratios}. The differing velocities in the different lines give strong constraints for the line ratios. Experiments with our model, for example, shows that it is the weakness of \OIII\ in the Cloud and nuclear disk that causes the dip in velocity of the Eastern part of the rotation curve to be deeper in \OIII\ and the \OIII\ rotation curve to be flatter on both sides. The main ambiguity with the model presented here is the difference in systemic velocity of the bulge and outer disk relative to the nucleus and nuclear emission line disk, as well as the relation of the Cloud to the emission line disk and to the bulge. The Cloud, although introduced in a somewhat ad hoc fashion, could be a component that resides in the emission line disk. Of course, to get all necessary information about the velocities and spatial distribution of the different components from an isolated spectrum is futile, because i.a. of the unknown influence due to the point spread function from sources outside the slit. By adding more slits, there is still a difficulty to obtain an accurate and complete coverage (Lindblad et al. 1996). Obviously, the ideal is an integral field spectrometer and an extension of the method to a two-dimensional treatment covering the entire region of interest.

\end{document}